\documentclass{template/iosart2c}
\usepackage{cite}
\usepackage{amsmath,amssymb,amsfonts}
\usepackage{algorithmic}
\usepackage{graphicx}
\usepackage{textcomp}
\usepackage[colorlinks]{hyperref}
\usepackage[table,xcdraw,dvipsnames]{xcolor}
\usepackage{multirow}
\usepackage{subcaption}
\usepackage{booktabs}
\usepackage{todonotes}
\usepackage{color}
\newcolumntype{d}[1]{D{.}{.}{#1}}

\firstpage{1} \lastpage{5} \volume{1} \pubyear{2021}

\begin{document}
\begin{frontmatter}

\title{RLupus}

\subtitle{Cooperation through emergent communication in \emph{The Werewolf} social deduction game}

\author[A]{\fnms{Nicolo'} \snm{Brandizzi}},
\author[B]{\fnms{Davide} \snm{Grossi}},
\author[A]{\fnms{Luca} \snm{Iocchi}}

\runningauthor{N. Brandizzi et al.}
\runningtitle{RLupus}

\address[A]{Dipartimento di Ingegneria Informatica, Automatica e Gestionale, \\ 
Sapienza University of Rome, Italy\\
E-mail: \{brandizzi, iocchi\}@diag.uniroma1.it}
\address[B]{Bernoulli Institute for Mathematics, Computer Science and Artificial Intelligence,\\ University of Groningen, The Netherlands. \\
Amsterdam Center for Law and Economics and Institute for Logic, Language and Computation, University of Amsterdam, The Netherlands.\\
E-mail: d.grossi@rug.nl}

\begin{abstract}
This paper focuses on the emergence of communication to support cooperation in environments modeled as social deduction games (SDG), that are games where players communicate freely to deduce each others' hidden intentions.
We first state the problem by giving a general formalization of SDG and a possible solution framework based on reinforcement learning.
Next, we focus on a specific SDG, known as \emph{The Werewolf}, and study if and how various forms of communication influence the outcome of the game.
Experimental results show that introducing a communication signal greatly increases the winning chances of a class of players. We also study the effect of the signal's length and range on the overall performance showing a non-linear relationship.

\end{abstract}


\end{frontmatter}

\onecolumn

\twocolumn
\section{Introduction}

Social deduction games (SGDs) are games characterized by the confrontation between two or more parties, one of which is usually seen as an evil faction. The other parties must deduce the real intention of the latter, seeing through lies and deceptions. 
While the details of these games may change, free communication---i.e., players can communicate with each other with no limitations---is a common aspect for them all.

In artificial settings, communication would lead to increased complexity in the environment, both on the user side, where engineers are tasked to design an expressive and robust syntax \cite{genesereth1992knowledge,finin1994kqml,o1998fipa}, and on the artificial players who have to learn the syntax and the meaning of the available words. For shallow players, this task quickly becomes unfeasible. A common solution is to define a game-specific language containing communication semantics. This language is devised by usually injecting some expert knowledge into the system. The language becomes then part of the game, providing new available communication actions augmenting the agents' choices in the decision-making process.
Instead of developing a language for the agents, in this article, we propose another approach enabling players to use a free communication mechanism. 

\paragraph{Context of the paper}
Games arising from social interaction have been extensively studied within the multi-agent reinforcement learning (MARL) literature. 
Indeed, MARL systems have been successfully used to model a wide variety of social systems found in nature and modern society, from food-collective ants systems to complex sharing of information in social networks \cite{bucsoniu2010multi}. In these settings, the agents' behavior can be cooperative, competitive or a mix of the two. In \cite{bucsoniu2010multi} a study of these different kinds of settings and the algorithms associated with them is pursued while \cite{tan1993multi,liang2020emergent} focuses on the difference between competitive and cooperative agents.

Although MARL has been used for a wide variety of applications, the reinforcement learning paradigm may not scale up easily in complex multi-agent systems. For this reason, RL has been integrated with deep neural networks (DeepRL, DRL) \cite{schmidhuber2015deep}. This allows RL to scale to problems that were previously intractable, such as playing video games using pixels as input \cite{mnih2013playing}. Shortly after \cite{tampuu2017multiagent}, this approach has been applied to multi-agent systems to study the complex emergent behavior of multiple agents interacting with each other to reach a goal. Although DRL is a well-established field of study, agent interaction via actions or communication remains a challenging problem.

This research area is strictly tied to the study of complex behavior arising from the interaction of simple agents; these aspects are mainly studied through the usage of structured game systems. In their work, Baker et al. \cite{baker2019emergent} showed how a few simple rules from the \emph{Hide'n Seek} game can generate complicated behaviors to the point of exploiting environmental errors to their advantages. Along the same line, Leibo et al. \cite{leibo2019autocurricula} carried out extensive analysis on the problem of autocurricula and non-stationary learning in multi-agent deep reinforcement learning (MADRL); they pointed out how the interaction of competitive agents can culminate in an endless cycle of counter-strategies due to the non-stationarity of the environment.

In particular, one such complex behavior consists in the emergence of communication between cooperative agents. Recent works investigate this aspect in various environments varying from joined image captioning \cite{graesser2019emergent}, to negotiation \cite{cao2018emergent} and simulated pointing games \cite{mordatch2018emergence}. Our paper is a contribution to this line of research, focusing on communication in social deduction games.

\paragraph{Paper contribution and outline}
The paper makes two main contributions:
\begin{itemize}
    \item a formal description of social deduction games coupled with a general reinforcement learning solution framework, which allows for free communication among agents without requiring to provide game-specific knowledge;
    \item an analysis of the performance obtained through learned communication behaviors in an instance of the above class of games: \emph{The Werewolf}\footnote{Also known as \emph{Lupus in Fabula}.} social deduction game.
\end{itemize}

The paper is organized as follows. An overview of the related work is provided in Section \ref{sec:related_word}. In Section \ref{sec:problem_stat}, the problem statement is formalized together with the general RL framework. Section \ref{sec:methods} describes the \emph{Werewolf} game instance in detail, defining both the game logic and the actual implementation. This game provides a fit ground to study language emergence since its whole system is based on communication, indeed it is the subject of an annual  \href{http://aiwolf.org/en/}{AiWolf} context in Japan.
The experimental settings and results are reported and commented in Section \ref{sec:results} together with the comparison between our work and \cite{aiwolf}.
Finally, a discussion is provided in Section \ref{sec:conclusion}.

The code is available at \url{https://github.com/nicofirst1/rl_werewolf}.
\section{Related Work}
\label{sec:related_word}

Our work relies on the findings coming from the social interaction field of psychology, coupled with the multi-agent systems and reinforcement learning.

\paragraph{Social Deduction Games.}
Social deduction games have been studied in the broader context of social interactions \cite{eger2018keeping}. In particular, they have been used to study the role of rationality in inter-personal interaction \cite{colman2003cooperation}, analyze the different forms of social mechanics \cite{consalvo2011using}, and research the role of communital topology \cite{abramson2001social}, however, the deduction part of these games has been neglected.

Indeed, in their work \cite{chan2009mathematical}, the authors give a mathematical formulation for a general social game in order to simplify the way to design such games; however, no specific formulation is given for deduction games. 

On the other hand, \cite{wiseman2019data} study the most influential information source in social deduction games and concludes that the interactions that occurred prior to the game are regarded as most important to the player. Although this approach is reasonable in the context of acquaintances, no result is given for games in which the playing parties do not know each other.

In all these works, the goal is centered around the social interaction between players. In our work, we shift the focus to finding an optimal policy to improve the performances of a party.

\paragraph{Multi-agent Deep RL}
Applying RL paradigms to find an optimal policy for multi-agent systems is a well-established line of work \cite{busoniu2008comprehensive,tan1993multi,shoham2003multi}.
In recent years deep neural networks (DNN) have been used to solve complex tasks such as playing Atari games \cite{mnih2013playing}, cooperating in Hide-and-seek \cite{baker2019emergent} and competing with humans on strategic games \cite{vinyals2019grandmaster,silver2016mastering}.

A common paradigm for taring RL agents in a deep setting is policy gradient methods, \cite{sutton1999policy,silver2014deterministic} where the gradient of a parametrized policy is used to guide the agent in the direction of maximal expected reward. In our work, we leverage a particular instance of these methods called Proximal Policy Optimization \cite{schulman2017proximal}.

In all such cases, the complexity of the environment coupled with a DNN generated unexpected behaviors that are usually counter-intuitive for humans but achieve greater performances in the task at hand. In our work, we leverage this aspect and further study how the coordination between agents varies under different communication instances.

\paragraph{Emergent Communication.}
This phenomenon has been exploited in the newly born field of emergent communication \cite{wagner2003progress}, where agents are given the choice to use a communication channel to achieve a common goal.  

The workshop on Emergent Communication (Emecom) includes many publications in the field of natural language processing \cite{li2020emergent} strictly tied to MADRL \cite{lazaridou2020emergent,liang2020emergent,foerster2016learning} and social deduction games as an environment.
Standard games for this line of research are the Task \& Talk \cite{kottur2017natural}, which is centered around dialogue, on the other hand, the Pointing Game \cite{mordatch2018emergence} grounds the communication into natural image processing.

Another instance of these settings is \emph{The Werewolf} game, where the players find themselves split into two opposite groups in a partially known environment. This game has gained increased popularity in the field of cooperation through emergent communication, especially in Japan, where the annual \hyperlink{http://aiwolf.org/en/}{AiWolf contest} \cite{bi2016human,nakamura2016constructing,hirata2016werewolf,katagami2014investigation} sees artificial agents competing with and against human players to win the game with fixed language syntax.
In particular, \cite{8588472} set up a 5-player game with additional roles and use a Deep Q-Network to determine who to trust or kill.

In our work, we choose \emph{The Werewolf} as an instance of the general SDG framework. However, our implementation differs from the ones in the \emph{AiWolf contest}, the closest being \cite{aiwolf}, where the authors use Q-learning to study the winning chances of the villagers in a game with 16 players, divided into 14 villagers and 2 werewolves. Indeed we drop the hand-coded syntax and let the players develop their own communication by defining some general attributes of the channel.

\section{Problem Statement}
\label{sec:problem_stat}

Social deduction games (SDG) are characterized by the presence of a number of opposing parties $$P^{(1)}, P^{(2)},\ldots, P^{(m)}$$ (typically $m=2$), each containing a finite number $n(k)$ of players, which may differ per party. For the sake of conciseness, we drop the $k$ dependency and denote the number of players in a party as simply $n$:

$$P^{(k)} = \{p^{(k)}_1,p^{(k)}_2,\ldots, p^{(k)}_n\}$$
with $1 \leq k \leq m$. We denote by $N = \bigcup_{1 \leq k \leq m} P^{(k)}$ the set of all players.

The game evolves as a sequence of actions performed by the players of the parties, typically in turns. The effect of these actions contributes to the definition of the game score.

The goal of each party is to prevail over the others by performing suitable social behaviors, including, for example, leveraging other players by means of bluffs and lies. These deceitful methods are the base for any SDG and force every player to perform a deductive analysis on the member of the other parties.
In general we can identify two types of goals in SDGs:
\begin{itemize}
    \item An agent-based \emph{micro-goal}, 
    which is the main factor steering the agent's behavior, either in isolation (in competitive environments) or together with other agents' goals (in cooperative environments).
    \item A party-based \emph{macro-goal}, expressing the aligned interests of the members of the same party that, combined, make up the party goal. 
\end{itemize}

During the execution of the game, agents can communicate among them, either implicitly or explicitly. We define explicit communication as the act of sharing information for the sole purpose of affecting other agents' mental states. On the other hand, implicit communication regards all those actions that carry more than one meaning, e.g., guiding someone to a goal.
Decisions about how, when, and what to communicate are critical choices for the game's success.

We thus distinguish two categories of actions performed by the players: 1) \emph{game actions}, that are actions affecting the evolution of the game, 2) 
\emph{communication actions} that are actions affecting only the mental state (i.e., the knowledge state) of the players.
In this formalization, we consider only forms of explicit communication, while studying forms of implicit communication is left as future work.

With the previous assumptions, in this paper, we consider the formalization of an SGD with the following elements:
\begin{enumerate}

    \item the action set $\mathcal{A}= \mathcal{G} \cup \mathcal{C}$ is made of two separate components:
    
    \begin{itemize}
         \item a finite set of possible game actions $\mathcal{G}$ the elements of which we will denote with $g$.
         \item a set of unidirectional communication actions $C_{i,j}(b)$ intended to convey some information $b \in \mathcal{B}$ \footnote{A possibly infinite set of all possible signals.} between two players:
         $$C_{i,j}(b): p_j \to p_i\quad \forall p_j,p_i \quad j\ne i \in N$$

    \end{itemize}
   
     \item the state set $\mathcal{O}=  \mathcal{E}^N \times \mathcal{W}^N \times \mathcal{V} $, with $N$ being the total number of players, built out of three elements\footnote{We consider a setup where all the agents $N$ choose an action simultaneously.}:
     
     \begin{itemize}
     
        \item a set of agent's features $\mathcal{E}$ representing the game situation of each agent (typically visible to all other agents),
     
        \item a set of agent's internal states $\mathcal{W}$ (e.g., representations of beliefs not visible to other agents)  .
        
        \item a set of environment states $\mathcal{V}$ that are common for all the agents (i.e., independent from the agent states).
     \end{itemize}

    \item an environment $\mathcal{S}$ implementing the game logic. $\mathcal{S}$ can be seen as a function taking as input the agents actions and yielding a new state obtained as the result of execution of such actions
    \begin{equation} \label{eq:trans}
        \mathcal{S}: \mathcal{O}\times \mathcal{A}^N \to \mathcal{O}
    \end{equation}

\end{enumerate}

Notice that, while it may be relatively easy to formalize the specifications of game actions, for example, in terms of pre-conditions and post-conditions using action representation formalisms, it is less clear how to formalize communication actions since it would require an explicit model of agents' knowledge. For modeling this kind of communication actions, the use of typical action formalisms is not straightforward.
For example, they may need to be extended with epistemic operators \cite{fagin95reasoning}.

\section{General solution framework}
\label{sec:framework}

This article studies social deduction games that can be formalized as a multi-agent (deep) reinforcement learning (RL) scenario. In such scenarios, each party has to choose an optimal strategy or policy \footnote{Although the coordinated behavior favors the parties' macro-goal, the actual policy works on the players' micro-goal level. } (i.e., an optimal assignment from states to actions) to maximize the game score. As already mentioned, a particular feature of SDGs is the presence of communication actions and the need to choose optimal communications among the players within a party.

The problem of learning optimal policies in multi-agent games is indeed well known, and many solutions are available. However, when communication actions are involved, using artificial intelligence techniques to make optimal decisions about how, what, and when to communicate is still a challenging problem under investigation, and fewer research works are available.

The advantage in defining a solution based on RL is that it does not require an explicit model of the transition function for communication actions. In other words, optimal behaviors can be computed without associating a semantic meaning to the communication actions.
While this feature can be considered not desirable for some kinds of applications (e.g., for mixed human-AI teams), it is very convenient for AI teams based on RL that can learn their own communication language to win the game. Indeed AI agents can effectively learn a communication language without making the semantics of communication explicit. The explainability of learned communication actions is left as future work.

\paragraph{Action Policy}
\label{sec:framework:policy}
We consider a single game turn $t$ as a sequence of $k+1$ steps, where the first $k$ steps are associated to only communication actions $c \in \mathcal{C}$ while the last step is a game action $g \in \mathcal{G}$. We denote each time step as
$\mathcal{O}_t^j \quad \forall j \in [0,k+1]$.

Now, we can define a policy $\pi$ that uses the information from $\mathcal{O}_t^j$ in order to choose an action as follows:
\begin{align*}
\pi(\mathcal{O}_t^j) \in \mathcal{C} \quad \forall \; j\le k \\
\pi(\mathcal{O}_t^{k+1}) \in \mathcal{G} \quad \,
\end{align*}
    
Note that in some cases it can be useful to compute two distinct policies $\pi_C\;; \pi_G$, one for communication and one for the game actions.

\section{R-Lupus framework}
\label{sec:methods}

In this section, we present the \emph{RLupus} framework for the Werewolf social deduction game in which we apply the above formalization.
In this specific context, we aim to study if and how various forms of communication can influence the outcome of the game, in which only one party is able to learn while the other one has a fixed, hand-coded policy.

Under the same circumstances, the hypothesis is that the agents who can communicate will perform much better than those not allowed to exchange information. Moreover, we speculate that different communication settings will have diverse influences on the amount of coordination among the agents and the outcome of the game.

In the following sub-sections, we give a brief introduction on \emph{The Werewolf} game logic,
a description of the RL environment with all its components and the policies used for the players in the game.

\subsection{The Werewolf Game}
\label{sec:method:game}

Werewolf is a social deduction game modeling conflicts between two groups in a partially known environment.
In its easiest version, the game sees two groups ($M=2$), villagers $P^{(v)}$ and werewolves $P^{(w)}$ where $P^{(v)}> P^{(w)} +1$. The wolves know exactly the identity of each player, while the villagers are certain exclusively about their role and the number of werewolves. In an open setup, an additional moderator is needed to coordinate the players.\\
The game is divided into two phases: night and day, interleaving each other.\\

The game ends either when the villagers execute the last werewolf, or there is an even number of both roles. The latter case implies the wolves winning since the execution phase can be stalled, thus taking away the only possibility for villagers to kill the wolves.

\begin{table}[ht]
\centering
\caption{RLupus: Multi-channel metrics. \normalfont{The first column reports the type of \textbf{comm}(unication)channel regarding the \textbf{S}ignal\textbf{L}ength and the \textbf{S}ignal\textbf{R}ange. The next four show the metrics values for villagers winning rate, suicide rare, number of days elapsed and accordance rate}}
\label{tab:9p_results_policy}
\begin{tabular}{lllll}
\toprule
\textbf{Comm} & \textbf{Win Vil} & \textbf{Suicide} & \textbf{Days} & \textbf{Accord} \\ \midrule
\textbf{0SL} & 0.044 & 0.086 & 1.55 & 0.47 \\
\textbf{1SL-2SR} & 0.19 & 0.078 & 1.58 & 0.47 \\
\textbf{1SL-9SR} & 0.21 & 0.078 & 1.58 & 0.47 \\
\textbf{9SL-2SR} & \textbf{0.45} & \textbf{0.067} & \textbf{1.9} & \textbf{0.47} \\
\textbf{9SL-9SR} & 0.19 & 0.077 & 1.58 & 0.46 \\ \bottomrule
\end{tabular}

\end{table}

\subsection{Reinforcement Learning}
\label{sec:method:rl}
Before describing the \emph{RLupus} environment, a brief description of the Reinforcement Learning paradigm must be introduced.

In the canonical RL environment, the problem satisfies the Markov property\footnote{The future depends only on the current state and action.} so it can be formulated as a Markov decision process (MDP). 
An MDP is defined by a tuple of five elements $(\mathcal{O}, \mathcal{A}, \mathcal{T}, \mathcal{R}, \gamma)$ defined as follows:
\begin{itemize}
    \item A finite set of observations called the observation space $\mathcal{O}$
    \item A finite set of actions, the action space $\mathcal{A}$
    \item A transition model $\mathcal{T}_a(m_t, m_{t+1})$ defining the probability to transition from state $m_t$ to a new state $ m_{t+1}$ given action $a$.
     \item $\mathcal{R}(m_t, m_{t+1})$ the reward associated to the previous transition. 
\end{itemize}

At each timestep $t$, an RL agents receives a state $m_t$ from a dynamic environment $\mathcal{O}$.
The agent then selects an action $a_t$ from the action space $\mathcal{A}$ following a policy $\pi(a_t|m_t)$. The action is processed by $\mathcal{S}$, equation \ref{eq:trans}, which transitions to the next state $m_{t+1}$ and yields a reward $r_t$ to the agent. 
This feedback loop continues until the agent reaches a terminal state and restarts.

The agent aims to maximize the expected discounted reward given by:
$$R_t=\sum_{i=0}^\infty \gamma^i r_{t+i}$$

Where $\gamma \in (0,1]$ is the discount factor governing the importance of future rewards.

\subsection{RLupus Environment}
\label{sec:method:env}

Dealing with RL implies the presence of an action and state set; the latter are referred to as action space and observation space, which are presented in the following section\footnote{For the sake of conciseness some formulations are omitted from the body of the paper. The interested reader can refer to the Appendix \ref{app:main}.}.

\paragraph{Action Space}
As mentioned in Section \ref{sec:method:policy}, the Werewolf uses a policy $\pi(\mathcal{O}_t^j)$ in order to choose both the communication and game actions.
In a way, the agent can be seen as performing a game $g_t$ and communication $c_t$ action simultaneously. 

\begin{figure}[t]
\centering
    \includegraphics[scale=0.4,keepaspectratio]{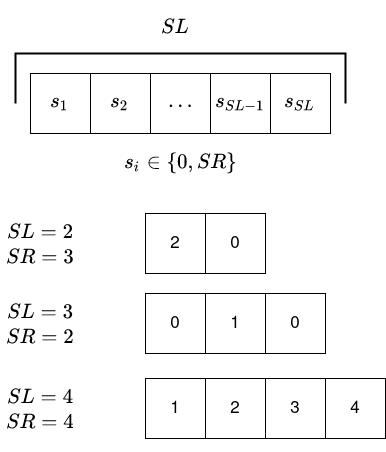}
    \caption{Signal vector with general formulation as well as specific values for \emph{SL} and \emph{SR}. }
    \label{fig:signal}
\end{figure}

Indeed the action space is divided into two parts: 
\begin{itemize}
    \item \emph{Target}: The target $g_t$ is an integer in range $g_t \in [0,N-1]$, where $N$ is the number of players. Its intended usage is to allow players to vote for other players during the game. The range of possible values never changes during the execution; instead, illegal actions, such as voting for dead players, are filtered out later in the model. 
    \item \emph{Signal}: The signal vector $c_t$ (see Figure \ref{fig:signal}) is defined by two integer values: its length $SL \in [0,\infty[ $ which can be any value from zero (no communication) to an arbitrary large integer; its range $SR \in [2,N]$ that defines the number of possible values it can have. $SR$ bounded below by $2$, since a signal with only one possible value would be considered a static vector carrying no information; the upper bound $N$ comes from the necessity for the signal to be embedded with the target. Both are used to define the valid space for communication before training.
\end{itemize}

\paragraph{Observation Space}
The observation space characterizes what the agents perceive in the environment. This space includes both other agents' actions and information about the environment\footnote{The complete observation space is described in Appendix \ref{app:obs_space}.}.

\paragraph{Transition Model}
The transition model fully depends on the target action; that is, the environment is independent of the communication between agents. Indeed $\mathcal{S}$ switches between night and day at every timestep $t$ and removes dead players from the environment.

\paragraph{Rewards}
The rewards, or penalties, are the core of the environment and determine how the players interact, learn, and develop new strategies; an agent's main goal is to take actions that will maximize the expected reward.\\
Following our formalization, the environment is responsible for delivering a reward to each player \footnote{The complete set of rewards is described in Appendix \ref{app:reward}.}.

\paragraph{Metrics}
To measure the changes in the agent behavior, the following normalized metrics are logged:

\begin{itemize}
    \item \emph{Suicide}: the number of times an agent votes for itself during an execution phase.
    
    \item \emph{Wins}: the villagers' wins are plotted in the normalized range of values.

    \item \emph{Average days} : average number of days before a match ends.

    \item \emph{Accord}:  This value represents, on average, the percentage of agents that vote for the same target during the two execution phases.

\end{itemize}

\begin{table*}[ht]
\centering
\caption{RLupus: single/no channel metrics. \normalfont{The Design Choice part of the table shows which kind of policy Random, Unite or Revenge has been used in relation to the communication, while the Results half present the metric' values}}
\label{tab:9p_results_signal}
\begin{tabular}{llll|llll}
\toprule
\multicolumn{4}{c|}{\textbf{Design Choice}} & \multicolumn{4}{c}{\textbf{Results}} \\ \hline
\textbf{Comm} & \textbf{Rnd} & \textbf{Unt} & \textbf{Rvg} & \textbf{Vil Win} & \textbf{Accord} & \textbf{Suicide} & \textbf{Days} \\ \midrule
\textbf{0SL} & X &  &  & 0.044 & 0.478 & 0.086 & 1.55 \\
\textbf{0SL} &  & X &  & 0.03 & \textbf{0.695} & \textbf{0.059} & \textbf{1.5} \\
\textbf{0SL} &  &  & X & \textbf{0.12} & 0.482 & 0.078 & 1.64 \\ \hline
\textbf{1SL-2SR} & X &  &  & 0.19 & 0.47 & 0.078 & 1.58 \\
\textbf{1SL-2SR} &  & X &  & 0.08 & \textbf{0.685} & \textbf{0.055} & \textbf{1.52} \\
\textbf{1SL-2SR} &  &  & X & \textbf{ 0.4} & 0.479 & 0.065 & 1.9 \\ \bottomrule
\end{tabular}

\end{table*}

\subsection{Policies}
\label{sec:method:policy}

An agent's policy defines the behavior of a player during the game. In this environment, there are two kinds of policies: \emph{trainable} policies use custom algorithms to collect experience and learn to maximize the reward; \emph{static} policies are hard-coded behaviors that are used to guarantee a fixed baseline trough out the evaluation.

In this work, we assign static policies to opponent players (werewolves) and training policies to AI agents (villagers) learning how to win the game.

\paragraph{Static policies for werewolves.}

Static policies are reserved for the werewolf agents; their aim is to allow a baseline evaluation of the villager learning. Since wolves are more likely to win in a completely random environment, applying such policies is enough to prove the development of new strategies for the villagers if the winning rates are to change significantly.\\
Three policy are implemented:
\begin{itemize}
    \item \emph{Random Target}: chooses a random non-dead player among the villagers during the execution phase. 
    \item \emph{Random Target Unite}: this policy targets the same player for every wolf, both during day and night execution; this allows the werewolf to dominate the day execution phase with random villagers.
    \item \emph{Revenge Target}: with this policy, the wolves will either vote randomly or target a villager who previously voted for a wolf.

\end{itemize}

\paragraph{Trainable policies for villagers.}
In our work, we chose Proximal Policy Optimization (PPO) as a training algorithm for its widespread success in multi-agent environment \cite{wei2019mixed,guan2020centralized}, but future research could also focus on other algorithms such as TRPO \cite{schulman2015trust} or MADDPG \cite{lowe2017multi}.
PPO uses a surrogate loss function to keep the difference between the old and the new policy within a safe range\footnote{More information on the policy loss are available in Appendix \ref{app:t_policy}.}.
The model is a simple fully connected network with an LSTM cell \footnote{Given the partial observability of the problem, we found the LSTM as a suitable element to approximate the state space.}.

\paragraph{Learning Werewolves.}
Up until now, only villager agents have been able to learn from experience, while the werewolf behave according to a static policy.
Here, we address the possibility of an environment where both villagers and werewolves are able to learn. 
Having multiple agents learning simultaneously void the stationary assumption that is necessary for optimality in RL. The presence of multiple villagers already invalidates this assumption, but the coordination between them alleviates the problem. Introducing an adversarial set of learning agents would cause an increased complexity that would cloud the goal of this paper, which is to study the emergence of a language between agents.

\section{Results}
\label{sec:results}

Following, an analysis of the results for both nine (Section \ref{sec:result:9p}) and twenty-one (Section \ref{sec:result:21p}) players is given.
In both cases, a baseline setting with no communication is compared with multi-channel communication to show the increased performance reported using the metrics in Section \ref{sec:method:env}.

Moreover, for the nine players instance, results for the additional \emph{revenge} and \emph{unite} policies are reported against the \emph{random} one.

Finally, in Section \ref{sec:result:16p}, a summary for a setting with 16 players is given for both the \emph{AiWolf} environment \cite{aiwolf} and the \emph{RLupus} one.

\subsection{Nine Players}
\label{sec:result:9p}
A game with nine players is relatively short but not trivial for the villagers. Indeed, in a completely random environment, they have a probability of $3.12\%$ to win\footnote{See Appendix \ref{app:baseline} for the complete estimation.}.

Since the random policy is believed to hold the least expert knowledge about the game environment, we present the results for this policy only in the next section. 
On the other hand, Section \ref{ssec:pol_comp} also shows the results for the \emph{revenge} and \emph{unite} policies.

\subsubsection{Multi-channel communication}

Table \ref{tab:9p_results_policy} presents the results (columns) obtained with different forms of communication (rows), with different communication settings (SL=signal length, SR = signal range).
From the table, it is clear that any form of communication improves over the non-communication setting (0SL). Moreover, it shows how the bit communication (2SR) performs much better than the full ranged one (9SR).
Indeed, with the addition of just one bit of communication, the villagers can perform as well as the full extended communication (9SL-9SR).
In fact, for the settings where $SR=9$, increasing the channel length $SL$ has the only effect of speeding up the convergence by a factor of circa $25\%$ for each increase.

On the other hand, the bit communication, independently from the channel length, provides generally better results. This leads to the conclusion that the two parameters dictating the change in the communication channel are not equally influential when it comes to the agent's learning. 

On a final note, the average time to train the agents was 6 hours on a single GPU\footnote{Nvidia Geforce GTX 1080 Ti.} machine.

\subsubsection{Policies comparison}
\label{ssec:pol_comp}

Table \ref{tab:9p_results_signal} reports results also related to the werewolves policies (Rnd = Random, Unt = Unite, Rvg = Revenge).

When no communication is available (0SL), the accord value is almost identical for both the revenge and the random policy and much higher for the unite one, as previously anticipated.

The number of days reports a similar trend being smaller in the unite settings where the game ends sooner.
Finally, the revenge winning rate reaches a much greater value than the other policies, $12\%$. The motivation being the simplicity of the policy itself, i.e., the wolves can not hide behind purely random action anymore and are not strong enough to drive the majority of the votes toward a villager.

Following the considerations for the three policies and bit communication ($SL=1\, SR=2$) referred to Table \ref{tab:9p_results_signal}:
\begin{itemize}
    \item \emph{Random}: the villager's winning rate reaches $20\%$ ;  $4.5$ times more than the previous condition and $6.5$ times the theoretical winning rate.  
    \item \emph{Unite}: as in the previous case, the coordination is much greater. Indeed the villagers are able to increase their winning rate by a factor of $\times 3$. This result alone proves that introducing a limited communication channel can greatly favor the outcome even in the most disadvantageous setting.
    \item \emph{Revenge}: again, the revenge policy is the easiest to spot for a trained agent reaching a winning rate of $40\%$.

\end{itemize}

\subsection{Twenty-one players}
\label{sec:result:21p}

Unlike what was studied in the previous section, the twenty-one players (21P) environment has more room for the villagers to win in a completely random setting. Indeed, by building a tabular representation of the expanded tree (shown in Table \ref{tab:results:21P:outcomes}), the reader can see how the total villagers' winning possibilities increased to $11.62\%$ in a completely random environment.
\begin{table}[h]
\centering
\caption{Mapping between outcomes and probabilities.}
\label{tab:results:21P:outcomes}
\begin{tabular}{cccc}
\toprule
\textbf{Outcome} & \textbf{Prob. \%} & \textbf{Leaves}         & { \textbf{Total \%}}     \\ \midrule
0-12             & 0.029             & \multicolumn{1}{c|}{1}  & { }                        \\
0-10             & 0.143             & \multicolumn{1}{c|}{4}  & { }                        \\
0-8              & 0.447             & \multicolumn{1}{c|}{10} & { }                        \\
0-6              & 1.162             & \multicolumn{1}{c|}{20} & { }                        \\
0-4              & 2.819             & \multicolumn{1}{c|}{35} & { }                        \\
0-2              & 7.02              & \multicolumn{1}{c|}{56} & \multirow{-6}{*}{{ 11.62}} \\\hline
1-1              & 21.06             & \multicolumn{1}{c|}{56} & { }                        \\
2-2              & 27.965            & \multicolumn{1}{c|}{21} & { }                        \\
3-3              & 26.017            & \multicolumn{1}{c|}{6}  & { }                        \\
4-4              & 26.017            & \multicolumn{1}{c|}{1}  & \multirow{-4}{*}{{ 88.38}} \\ \hline

\textbf{Total}   & 1.0               & 210                     & { 1.0}                       \\ 
\bottomrule
\end{tabular}
\end{table}

\subsubsection{Multi channel communication}

A comparison among all the multi-channel settings is reported in Table \ref{tab:21p_results}.

According to the previous section's findings, the bit communication ($9SL-2SR$) seems to perform better than the full-ranged one ($*SL-21SR$) both in terms of winning ratio and suicides.

On the other hand, the $1SL-2SR$ instance performs worse than the no communication one; this anomaly can be attributed to an incorrect environment exploration. As mentioned in Table \ref{tab:results:21P:outcomes}, the expanded tree size is greater than the instance with nine players; thus, the algorithm can get more easily stuck in a local minimum.
However, every other setting with a communication channel has a higher winning rate than the one without; thus, one could confidently say that the communication signal is indeed helping the agents exploring the possible branches better than in other cases.
For these reasons, we believe this outcome shows how the shape of the communication signal can improve the agents' capacity to explore the environment and thus should be further investigated in future work.

\begin{table}[ht]
\centering
\caption{RLupus: 21 Player metrics.}
\label{tab:21p_results}
\begin{tabular}{lllll}
\toprule
\textbf{Comm} & \textbf{Win} & \textbf{Suicide} & \textbf{Days} & \textbf{Accord} \\ \midrule
\textbf{0SL} & 0.42 & 0.072 & 7.84 & 0.56 \\
\textbf{1SL-2SR} & 0.25 & 0.075 & 7.74 & \textbf{0.57} \\
\textbf{9SL-2SR} & \textbf{0.98} & \textbf{0.04} &\textbf{ 7.3} & 0.56 \\
\textbf{21SL-2SR} & 0.94 & 0.05 & 7.6 & 0.56 \\
\textbf{1SL-21SR} & 0.72 & 0.062 & 8 & 0.56 \\
\textbf{21SL-21SR} & 0.61 & 0.066 & 7.9 & 0.56 \\ \bottomrule
\end{tabular}

\end{table}

\subsection{Sixteen Players}
\label{sec:result:16p}

As previously mentioned, Kajiwara et al. \cite{aiwolf} presented the AiWolf framework,  implementing a well-defined semantic\footnote{More details about the AiWolf framework are available in Appendix \ref{app:aiwolf}.} constrained to be both easy to use for the artificial players and understandable by humans. They extended the agents' adaptive capabilities by adopting a Deep Q-network architecture and reported an increase in the villagers' winning rate when they are the only ones capable of learning.
Although their methodology is similar to ours, their work is not directly comparable to RLupus because the two frameworks represent the game in different ways. The different set-ups lead to related but orthogonal findings, which we discuss in this section.

We trained an environment with 16 players (2 werewolves and 14 villagers) and reported our results in Table \ref{tab:p16_comparison}, together with the results of the AiWolf framework.

\begin{table}[h]
\centering
\caption{Winning rate for \cite{aiwolf} and our method in both baseline and active learning.}
\label{tab:p16_comparison}
\begin{tabular}{c|l|cc}
\toprule
\multicolumn{2}{c|}{\multirow{2}{*}{\textbf{Design}}} & \multicolumn{2}{c}{\textbf{Win Vil}} \\ \cline{3-4} 
\multicolumn{2}{c|}{} & AiWolf & RLupus \\ \hline
\multirow{2}{*}{No Learning} & AiWolf baseline & 38.6 & / \\
 & RLupus baseline & / & 56.9 \\ \hline
\multirow{3}{*}{Learning} & DQN / hand-coded & 52.9 & / \\
 & PPO / 0SL & / & 90.2 \\
 & PPO / 1SL-2SR & / & 98.4 \\ \bottomrule
\end{tabular}
\end{table}

It is interesting to observe the increase in performance due to the learning process.
Indeed, the table shows baseline results without learning in the two frameworks and results obtained with RL.

In comparison with AiWolf, 
it can be seen how the 0SL setting in RLupus already improves the winning chances by $33\%$ reaching $90.2\%$ and adding a single bit communication results in almost perfect victory on the villagers' side ($98.4\%$).

\section{Conclusions and Future Work}
\label{sec:conclusion}

In this paper, we have defined a formalism for social deduction games, in which communication is an essential part of the game together with the allowed actions.
On top of that, a general resolution framework based on reinforcement learning was defined and applied to the \emph{The Werewolf} SDG by studying how various forms of communication influenced the outcomes of a match.

As shown by the experimental results, the introduction of different forms of communication greatly increases the agents' performance (villagers). 
In particular, we observed that a Boolean signal range is preferred to an integer one. The reason for this is unclear. We speculate it may lay in the duality of the roles of the game.

Moreover, we found how much of the villagers' winning rate is determined by the agents not voting for themselves while keeping the accord value to a maximum. This translates into better coordination, which is possible only when there is a sufficient amount of communication present in the environment.

Also, we noticed that there is no linear map between the amount of communication permitted, i.e., \emph{SL} and \emph{SR}, and the overall performance. Indeed there seems to be an optimal combination of the two, which depends mainly on the signal length.


\medskip

We conclude by identifying three directions in which our work could be extended.

\paragraph{Model of SDGs}
One possible line of research consists of studying an instance of an SDG where the communication cannot be intrinsically tied to the action space, i.e., multi-step communication games such as Task \& Talk \cite{kottur2017natural}.
Alternatively, one could decide to depart from the deep part of the RL resolution framework and choose an SDG instance whose environment can be optically solved with standard reinforcement methods, e.g., Pointing game \cite{mordatch2018emergence}.

\paragraph{The Werewolf}
On the other hand, various possibilities arise from the study of The Werewolf game. 
One such could be the analysis of the language used for communication to highlight potential patterns, given that performance alone should not be considered a valid metric for the study of emergent communication \cite{lowe2019pitfalls}.
Moreover, one could extend the RLupus environment either by adding other roles, for example, the medium and the witch, or using normalized continuous vectors for the communication channel, which allow for the backpropagation of errors directly through the communication channel \cite{foerster2016learning} \footnote{On the other hand, \cite{jang2016categorical} and \cite{maddison2016concrete} use a Gumbel approximation to backpropagate the error through a discrete distribution. }.
Finally, a deeper analysis could be applied to understand the impact of the communication parameters (SR and SL) on the metrics of the system and the overall performance. 

\paragraph{Human-machine coordination}
Coordination among artificial agents is a key engineering challenge: from task allocation \cite{vytelingum2010agent,de2012multiagent}, knowledge management \cite{wu2001software}, distributed constraint optimization problems \cite{modi2005adopt} to multi-robot SLAM \cite{zhou2006multi,thrun2005multi} and language models \cite{bengio2003neural,vaswani2017attention}. And even more challenging is coordination between artificial and human agents. The study of emergent communication in SDGs could provide useful novel insights for the development of more efficient coordination among artificial agents and between artificial and human agents.

\bibliographystyle{apalike}
\bibliography{citations}

\begin{thebibliography}{}

\bibitem[Abramson and Kuperman, 2001]{abramson2001social}
Abramson, G. and Kuperman, M. (2001).
\newblock Social games in a social network.
\newblock {\em Physical Review E}, 63(3):030901.

\bibitem[Ahmed et~al., 2019]{ahmed2018understanding}
Ahmed, Z., Roux, N.~L., Norouzi, M., and Schuurmans, D. (2019).
\newblock Understanding the impact of entropy on policy optimization.
\newblock In Chaudhuri, K. and Salakhutdinov, R., editors, {\em Proceedings of
  the 36th International Conference on Machine Learning, {ICML} 2019, 9-15 June
  2019, Long Beach, California, {USA}}, volume~97 of {\em Proceedings of
  Machine Learning Research}, pages 151--160. {PMLR}.

\bibitem[Baker et~al., 2020]{baker2019emergent}
Baker, B., Kanitscheider, I., Markov, T.~M., Wu, Y., Powell, G., McGrew, B.,
  and Mordatch, I. (2020).
\newblock Emergent tool use from multi-agent autocurricula.
\newblock In {\em 8th International Conference on Learning Representations,
  {ICLR} 2020, Addis Ababa, Ethiopia, April 26-30, 2020}. OpenReview.net.

\bibitem[Bengio et~al., 2003]{bengio2003neural}
Bengio, Y., Ducharme, R., Vincent, P., and Janvin, C. (2003).
\newblock A neural probabilistic language model.
\newblock {\em The journal of machine learning research}, 3:1137--1155.

\bibitem[Bi and Tanaka, 2016]{bi2016human}
Bi, X. and Tanaka, T. (2016).
\newblock Human-side strategies in the werewolf game against the stealth
  werewolf strategy.
\newblock In {\em International Conference on Computers and Games}, pages
  93--102. Springer.

\bibitem[Busoniu et~al., 2008]{busoniu2008comprehensive}
Busoniu, L., Babuska, R., and De~Schutter, B. (2008).
\newblock A comprehensive survey of multiagent reinforcement learning.
\newblock {\em IEEE Transactions on Systems, Man, and Cybernetics, Part C
  (Applications and Reviews)}, 38(2):156--172.

\bibitem[Bu{\c{s}}oniu et~al., 2010]{bucsoniu2010multi}
Bu{\c{s}}oniu, L., Babu{\v{s}}ka, R., and De~Schutter, B. (2010).
\newblock Multi-agent reinforcement learning: An overview.
\newblock In {\em Innovations in multi-agent systems and applications-1}, pages
  183--221. Springer.

\bibitem[Cao et~al., 2018]{cao2018emergent}
Cao, K., Lazaridou, A., Lanctot, M., Leibo, J.~Z., Tuyls, K., and Clark, S.
  (2018).
\newblock Emergent communication through negotiation.
\newblock In {\em 6th International Conference on Learning Representations,
  {ICLR} 2018, Vancouver, BC, Canada, April 30 - May 3, 2018, Conference Track
  Proceedings}. OpenReview.net.

\bibitem[Chan et~al., 2009]{chan2009mathematical}
Chan, K.~T., King, I., and Yuen, M.-C. (2009).
\newblock Mathematical modeling of social games.
\newblock In {\em 2009 International Conference on Computational Science and
  Engineering}, volume~4, pages 1205--1210. IEEE.

\bibitem[Colman, 2003]{colman2003cooperation}
Colman, A.~M. (2003).
\newblock Cooperation, psychological game theory, and limitations of
  rationality in social interaction.
\newblock {\em Behavioral and brain sciences}, 26(2):139--153.

\bibitem[Consalvo, 2011]{consalvo2011using}
Consalvo, M. (2011).
\newblock Using your friends: Social mechanics in social games.
\newblock In {\em Proceedings of the 6th International Conference on
  Foundations of Digital Games}, pages 188--195.

\bibitem[de~Weerdt et~al., 2012]{de2012multiagent}
de~Weerdt, M.~M., Zhang, Y., and Klos, T. (2012).
\newblock Multiagent task allocation in social networks.
\newblock {\em Autonomous Agents and Multi-Agent Systems}, 25(1):46--86.

\bibitem[Eger and Martens, 2018]{eger2018keeping}
Eger, M. and Martens, C. (2018).
\newblock Keeping the story straight: A comparison of commitment strategies for
  a social deduction game.
\newblock In {\em Fourteenth Artificial Intelligence and Interactive Digital
  Entertainment Conference}.

\bibitem[Fagin et~al., 1995]{fagin95reasoning}
Fagin, R., Halpern, J.~Y., Moses, Y., and Vardi, M.~Y. (1995).
\newblock {\em Reasoning About Knowledge}.
\newblock {MIT} Press.

\bibitem[Finin et~al., 1994]{finin1994kqml}
Finin, T., Fritzson, R., McKay, D., and McEntire, R. (1994).
\newblock Kqml as an agent communication language.
\newblock In {\em Proceedings of the third international conference on
  Information and knowledge management}, pages 456--463.

\bibitem[Foerster et~al., 2016]{foerster2016learning}
Foerster, J., Assael, I.~A., De~Freitas, N., and Whiteson, S. (2016).
\newblock Learning to communicate with deep multi-agent reinforcement learning.
\newblock In {\em Advances in neural information processing systems}, pages
  2137--2145.

\bibitem[Genesereth et~al., 1992]{genesereth1992knowledge}
Genesereth, M.~R., Fikes, R.~E., et~al. (1992).
\newblock Knowledge interchange format-version 3.0: reference manual.

\bibitem[Graesser et~al., 2019]{graesser2019emergent}
Graesser, L., Cho, K., and Kiela, D. (2019).
\newblock Emergent linguistic phenomena in multi-agent communication games.
\newblock In Inui, K., Jiang, J., Ng, V., and Wan, X., editors, {\em
  Proceedings of the 2019 Conference on Empirical Methods in Natural Language
  Processing and the 9th International Joint Conference on Natural Language
  Processing, {EMNLP-IJCNLP} 2019, Hong Kong, China, November 3-7, 2019}, pages
  3698--3708. Association for Computational Linguistics.

\bibitem[Guan et~al., 2020]{guan2020centralized}
Guan, Y., Ren, Y., Li, S.~E., Sun, Q., Luo, L., and Li, K. (2020).
\newblock Centralized cooperation for connected and automated vehicles at
  intersections by proximal policy optimization.
\newblock {\em IEEE Transactions on Vehicular Technology}, 69(11):12597--12608.

\bibitem[Hirata et~al., 2016]{hirata2016werewolf}
Hirata, Y., Inaba, M., Takahashi, K., Toriumi, F., Osawa, H., Katagami, D., and
  Shinoda, K. (2016).
\newblock Werewolf game modeling using action probabilities based on play log
  analysis.
\newblock In {\em International Conference on Computers and Games}, pages
  103--114. Springer.

\bibitem[Hosu and Rebedea, 2016]{mnih2013playing}
Hosu, I. and Rebedea, T. (2016).
\newblock Playing atari games with deep reinforcement learning and human
  checkpoint replay.
\newblock {\em CoRR}, abs/1607.05077.

\bibitem[Jang et~al., 2017]{jang2016categorical}
Jang, E., Gu, S., and Poole, B. (2017).
\newblock Categorical reparameterization with gumbel-softmax.
\newblock In {\em 5th International Conference on Learning Representations,
  {ICLR} 2017, Toulon, France, April 24-26, 2017, Conference Track
  Proceedings}. OpenReview.net.

\bibitem[Kajiwara et~al., 2014]{aiwolf}
Kajiwara, K., Toriumi, F., Ohashi, H., Osawa, H., Katagami, D., Inaba, M.,
  Shinoda, K., Nishino, J., et~al. (2014).
\newblock Extraction of optimal strategies in human wolf using reinforcement
  learning.
\newblock {\em Proceedings of the 76th National Convention}, 2014(1):597--598.

\bibitem[Katagami et~al., 2014]{katagami2014investigation}
Katagami, D., Takaku, S., Inaba, M., Osawa, H., Shinoda, K., Nishino, J., and
  Toriumi, F. (2014).
\newblock Investigation of the effects of nonverbal information on werewolf.
\newblock In {\em 2014 IEEE International Conference on Fuzzy Systems
  (FUZZ-IEEE)}, pages 982--987. IEEE.

\bibitem[Kottur et~al., 2017]{kottur2017natural}
Kottur, S., Moura, J. M.~F., Lee, S., and Batra, D. (2017).
\newblock Natural language does not emerge 'naturally' in multi-agent dialog.
\newblock In Palmer, M., Hwa, R., and Riedel, S., editors, {\em Proceedings of
  the 2017 Conference on Empirical Methods in Natural Language Processing,
  {EMNLP} 2017, Copenhagen, Denmark, September 9-11, 2017}, pages 2962--2967.
  Association for Computational Linguistics.

\bibitem[Lazaridou and Baroni, 2020]{lazaridou2020emergent}
Lazaridou, A. and Baroni, M. (2020).
\newblock Emergent multi-agent communication in the deep learning era.
\newblock {\em CoRR}, abs/2006.02419.

\bibitem[Leibo et~al., 2019]{leibo2019autocurricula}
Leibo, J.~Z., Hughes, E., Lanctot, M., and Graepel, T. (2019).
\newblock Autocurricula and the emergence of innovation from social
  interaction: {A} manifesto for multi-agent intelligence research.
\newblock {\em CoRR}, abs/1903.00742.

\bibitem[Li et~al., 2020]{li2020emergent}
Li, Y., Ponti, E.~M., Vulic, I., and Korhonen, A. (2020).
\newblock Emergent communication pretraining for few-shot machine translation.
\newblock In Scott, D., Bel, N., and Zong, C., editors, {\em Proceedings of the
  28th International Conference on Computational Linguistics, {COLING} 2020,
  Barcelona, Spain (Online), December 8-13, 2020}, pages 4716--4731.
  International Committee on Computational Linguistics.

\bibitem[Liang et~al., 2020]{liang2020emergent}
Liang, P.~P., Chen, J., Salakhutdinov, R., Morency, L., and Kottur, S. (2020).
\newblock On emergent communication in competitive multi-agent teams.
\newblock In Seghrouchni, A. E.~F., Sukthankar, G., An, B., and Yorke{-}Smith,
  N., editors, {\em Proceedings of the 19th International Conference on
  Autonomous Agents and Multiagent Systems, {AAMAS} '20, Auckland, New Zealand,
  May 9-13, 2020}, pages 735--743. International Foundation for Autonomous
  Agents and Multiagent Systems.

\bibitem[Lowe et~al., 2019]{lowe2019pitfalls}
Lowe, R., Foerster, J.~N., Boureau, Y., Pineau, J., and Dauphin, Y.~N. (2019).
\newblock On the pitfalls of measuring emergent communication.
\newblock In Elkind, E., Veloso, M., Agmon, N., and Taylor, M.~E., editors,
  {\em Proceedings of the 18th International Conference on Autonomous Agents
  and MultiAgent Systems, {AAMAS} '19, Montreal, QC, Canada, May 13-17, 2019},
  pages 693--701. International Foundation for Autonomous Agents and Multiagent
  Systems.

\bibitem[Lowe et~al., 2017]{lowe2017multi}
Lowe, R., Wu, Y., Tamar, A., Harb, J., Abbeel, P., and Mordatch, I. (2017).
\newblock Multi-agent actor-critic for mixed cooperative-competitive
  environments.
\newblock {\em arXiv preprint arXiv:1706.02275}.

\bibitem[Luo et~al., 2018]{luo2018visual}
Luo, J., Green, S., Feghali, P., Legrady, G., and Ko{\c{c}}, {\c{C}}.~K.
  (2018).
\newblock Visual diagnostics for deep reinforcement learning policy
  development.
\newblock {\em CoRR}, abs/1809.06781.

\bibitem[Maddison et~al., 2017]{maddison2016concrete}
Maddison, C.~J., Mnih, A., and Teh, Y.~W. (2017).
\newblock The concrete distribution: {A} continuous relaxation of discrete
  random variables.
\newblock In {\em 5th International Conference on Learning Representations,
  {ICLR} 2017, Toulon, France, April 24-26, 2017, Conference Track
  Proceedings}. OpenReview.net.

\bibitem[Modi et~al., 2005]{modi2005adopt}
Modi, P.~J., Shen, W.-M., Tambe, M., and Yokoo, M. (2005).
\newblock Adopt: Asynchronous distributed constraint optimization with quality
  guarantees.
\newblock {\em Artificial Intelligence}, 161(1-2):149--180.

\bibitem[Mordatch and Abbeel, 2018]{mordatch2018emergence}
Mordatch, I. and Abbeel, P. (2018).
\newblock Emergence of grounded compositional language in multi-agent
  populations.
\newblock In {\em Thirty-Second AAAI Conference on Artificial Intelligence}.

\bibitem[Nakamura et~al., 2016]{nakamura2016constructing}
Nakamura, N., Inaba, M., Takahashi, K., Toriumi, F., Osawa, H., Katagami, D.,
  and Shinoda, K. (2016).
\newblock Constructing a human-like agent for the werewolf game using a
  psychological model based multiple perspectives.
\newblock In {\em 2016 IEEE Symposium Series on Computational Intelligence
  (SSCI)}, pages 1--8. IEEE.

\bibitem[O'Brien and Nicol, 1998]{o1998fipa}
O'Brien, P.~D. and Nicol, R.~C. (1998).
\newblock Fipa—towards a standard for software agents.
\newblock {\em BT Technology Journal}, 16(3):51--59.

\bibitem[Schmidhuber, 2015]{schmidhuber2015deep}
Schmidhuber, J. (2015).
\newblock Deep learning in neural networks: An overview.
\newblock {\em Neural networks}, 61:85--117.

\bibitem[Schulman et~al., 2015]{schulman2015trust}
Schulman, J., Levine, S., Abbeel, P., Jordan, M., and Moritz, P. (2015).
\newblock Trust region policy optimization.
\newblock In {\em International conference on machine learning}, pages
  1889--1897.

\bibitem[Schulman et~al., 2017]{schulman2017proximal}
Schulman, J., Wolski, F., Dhariwal, P., Radford, A., and Klimov, O. (2017).
\newblock Proximal policy optimization algorithms.
\newblock {\em CoRR}, abs/1707.06347.

\bibitem[Shoham et~al., 2003]{shoham2003multi}
Shoham, Y., Powers, R., and Grenager, T. (2003).
\newblock Multi-agent reinforcement learning: a critical survey.
\newblock {\em Web manuscript}, 2.

\bibitem[Silver et~al., 2016]{silver2016mastering}
Silver, D., Huang, A., Maddison, C.~J., Guez, A., Sifre, L., Van Den~Driessche,
  G., Schrittwieser, J., Antonoglou, I., Panneershelvam, V., Lanctot, M.,
  et~al. (2016).
\newblock Mastering the game of go with deep neural networks and tree search.
\newblock {\em nature}, 529(7587):484--489.

\bibitem[Silver et~al., 2014]{silver2014deterministic}
Silver, D., Lever, G., Heess, N., Degris, T., Wierstra, D., and Riedmiller, M.
  (2014).
\newblock Deterministic policy gradient algorithms.

\bibitem[Sutton et~al., 1999]{sutton1999policy}
Sutton, R.~S., McAllester, D., Singh, S., and Mansour, Y. (1999).
\newblock Policy gradient methods for reinforcement learning with function
  approximation.
\newblock {\em Advances in neural information processing systems},
  12:1057--1063.

\bibitem[Tampuu et~al., 2017]{tampuu2017multiagent}
Tampuu, A., Matiisen, T., Kodelja, D., Kuzovkin, I., Korjus, K., Aru, J., Aru,
  J., and Vicente, R. (2017).
\newblock Multiagent cooperation and competition with deep reinforcement
  learning.
\newblock {\em PloS one}, 12(4):e0172395.

\bibitem[Tan, 1993]{tan1993multi}
Tan, M. (1993).
\newblock Multi-agent reinforcement learning: Independent vs. cooperative
  agents.
\newblock In {\em Proceedings of the tenth international conference on machine
  learning}, pages 330--337.

\bibitem[Thrun and Liu, 2005]{thrun2005multi}
Thrun, S. and Liu, Y. (2005).
\newblock Multi-robot slam with sparse extended information filers.
\newblock In {\em Robotics Research. The Eleventh International Symposium},
  pages 254--266. Springer.

\bibitem[Vaswani et~al., 2017]{vaswani2017attention}
Vaswani, A., Shazeer, N., Parmar, N., Uszkoreit, J., Jones, L., Gomez, A.~N.,
  Kaiser, L., and Polosukhin, I. (2017).
\newblock Attention is all you need.
\newblock In Guyon, I., von Luxburg, U., Bengio, S., Wallach, H.~M., Fergus,
  R., Vishwanathan, S. V.~N., and Garnett, R., editors, {\em Advances in Neural
  Information Processing Systems 30: Annual Conference on Neural Information
  Processing Systems 2017, December 4-9, 2017, Long Beach, CA, {USA}}, pages
  5998--6008.

\bibitem[Vinyals et~al., 2019]{vinyals2019grandmaster}
Vinyals, O., Babuschkin, I., Czarnecki, W.~M., Mathieu, M., Dudzik, A., Chung,
  J., Choi, D.~H., Powell, R., Ewalds, T., Georgiev, P., et~al. (2019).
\newblock Grandmaster level in starcraft ii using multi-agent reinforcement
  learning.
\newblock {\em Nature}, 575(7782):350--354.

\bibitem[Vytelingum et~al., 2010]{vytelingum2010agent}
Vytelingum, P., Voice, T.~D., Ramchurn, S.~D., Rogers, A., and Jennings, N.~R.
  (2010).
\newblock Agent-based micro-storage management for the smart grid.

\bibitem[Wagner et~al., 2003]{wagner2003progress}
Wagner, K., Reggia, J.~A., Uriagereka, J., and Wilkinson, G.~S. (2003).
\newblock Progress in the simulation of emergent communication and language.
\newblock {\em Adaptive Behavior}, 11(1):37--69.

\bibitem[{Wang} and {Kaneko}, 2018]{8588472}
{Wang}, T. and {Kaneko}, T. (2018).
\newblock Application of deep reinforcement learning in werewolf game agents.
\newblock In {\em 2018 Conference on Technologies and Applications of
  Artificial Intelligence (TAAI)}, pages 28--33.

\bibitem[Wei et~al., 2019]{wei2019mixed}
Wei, H., Liu, X., Mashayekhy, L., and Decker, K. (2019).
\newblock Mixed-autonomy traffic control with proximal policy optimization.
\newblock In {\em 2019 IEEE Vehicular Networking Conference (VNC)}, pages 1--8.
  IEEE.

\bibitem[Wiseman and Lewis, 2019]{wiseman2019data}
Wiseman, S. and Lewis, K. (2019).
\newblock What data do players rely on in social deduction games?
\newblock In {\em Extended Abstracts of the Annual Symposium on Computer-Human
  Interaction in Play Companion Extended Abstracts}, pages 781--787.

\bibitem[Wu, 2001]{wu2001software}
Wu, D.-J. (2001).
\newblock Software agents for knowledge management: coordination in multi-agent
  supply chains and auctions.
\newblock {\em Expert Systems with Applications}, 20(1):51--64.

\bibitem[Zhou and Roumeliotis, 2006]{zhou2006multi}
Zhou, X.~S. and Roumeliotis, S.~I. (2006).
\newblock Multi-robot slam with unknown initial correspondence: The robot
  rendezvous case.
\newblock In {\em 2006 IEEE/RSJ international conference on intelligent robots
  and systems}, pages 1785--1792. IEEE.

\end{thebibliography}
\newpage
\onecolumn

\appendix

\section{Appendix}
\label{app:main}
Following, some specific aspects for the \emph{RLupus} environment are discussed. 
We decided to provide them in the Appendix to avoid interrupting the normal flow of the paper.
Indeed what comes in the next paragraphs are implementation details which are not required to understand the contribution of this article. 

\subsection{Environment}

\subsubsection{Observation Space}
\label{app:obs_space}
The observation space characterizes what the agents perceive in the environment. Multiple elements make up this space in the environment; each of them is later packed into a common dictionary and fed to the players:
\begin{itemize}
    \item \emph{Phase}: $p\in \{1,2,3,4\}$, to represent the four different phases of the game.
    \item \emph{Day}: $d\in \mathcal{I}^+$, to represent the day. An upper bound for the possible maximum day is set to $10$ which reduces the one-hot encoder representation length.
    \item \emph{Status Map}: $d\in \mathcal{B}$, this array maps a boolean value to an agent index, effectively informing the players if agent $i$ is alive $=1$ or dead $=0$. 
    \item \emph{Own id}: $oi \in [0,N]$, at the start of each game the agents' ids are shuffled. To keep an agent informed about its new position this integer is necessary.
    \item \emph{Targets}: $t \in [-1,N]$, to group all the targets.
    \item \emph{Signal}: $s \in [-1,SR-1]$, to group all the signals when there are any.
\end{itemize}

\subsubsection{Rewards}
\label{app:reward}

In the environment there are five conditions that determine if an agent is punished or rewarded:
\begin{itemize}
    \item \emph{Day}: At the end of each day every agent is penalized by a small factor  ($-1$). The purpose is to train the agents to quickly win the game, trying to develop new policies to win faster. 

    \item \emph{Death}: Each time a player dies, it takes a $-5$ penalty. While dying can be a strategy in some matches, this behavior needs to be controlled to avoid a high number of suicides.

    \item \emph{Target Accord}: To keep the voting system become another form of communication, the agents are penalized when they voted for someone that later on are not killed. Since execution depends on the coordination and communication of the required group of agents, this rewards is aimed to encourage a cooperative behavior
    
    \item \emph{Win/Lost}: Until now the rewards were assigned per agent. At the end of a match each group, either werewolves or villagers, is rewarded or penalized by a factor $\pm 25$.\\

\end{itemize}

\subsubsection{Trainable Policy}
\label{app:t_policy}

Since the neural network architecture is positioned in between the policy and the value function, the loss function must take into account the contribution given by the latter, thus an additional error term must be included:

$$
L^{CLIP+VF}=E_t\left [ L_t^{CLIP}(\theta) - c_1L_t^{VF}(\theta) \right]
$$

Where $L_t^{VF}(\theta) =(V_\theta(s_t)-V_t^{targ})^2$ is the squared-error loss between the value function $V_\theta$ at state $s_t$ and the target function, $c_1$ is its coefficient. This addiction is of fundamental importance to estimate the critic loss, that is the how well the model is able to predict the value of each state.\\

An additional term is introduced to regularize the total loss:
$$
L^{CLIP+VF+S}=E_t\left [ L_t^{CLIP+VS}(\theta) + c_2 S[\pi_\theta](s_t) \right]
$$
The entropy coefficient is maximized when all the policies have equal probability to be chosen, that is when the agent is acting at random. Adding such value to the loss function incentives the training algorithm to minimize the entropy value, thus avoiding the premature convergence of one action probability dominating the policy and preventing exploration \footnote{For a complete understanding of the influence entropy has on policy optimizations, the reader is referred to \cite{ahmed2018understanding}.}. The term $c_2$ scales the value of the entropy.

\vfill
\subsection{Baselines}
\label{app:baseline}
Understanding how AIs learn is a hard problem in the field of machine learning. In computer vision and specifically with convolutional neural networks (CNN), visualizing the activation layers together with the filters is a common practice to better understand the structure of the network itself.\\
On the other hand, for RL environments, there has been some research in the area of visual diagnostic for RL systems using CNNs to elaborate visual stimulus \cite{luo2018visual}; for emergent language, the preferred way to quantitatively evaluate the collaboration between agents has been accomplished by measuring the performances of the agents themselves.\\

It is trivial to see that, whichever method is preferred, there must be some kind of evaluation method for the communication system. In the following, such evaluation is estimated firstly by drawing some baselines with complete random agents, and then by comparing the latter with the results yield by the training.\\

A game with nine players is relatively short. \\
These agents determine a simple set of actions that can be easily mapped in a table; for this reason, the following section will focus only on nine players' settings (9P). Before showing any results the study of the complete behavior of a 9P setting is performed by inspecting its tree of possibilities.\\

\begin{figure}[h]
\centering
    \includegraphics[scale=0.4,keepaspectratio]{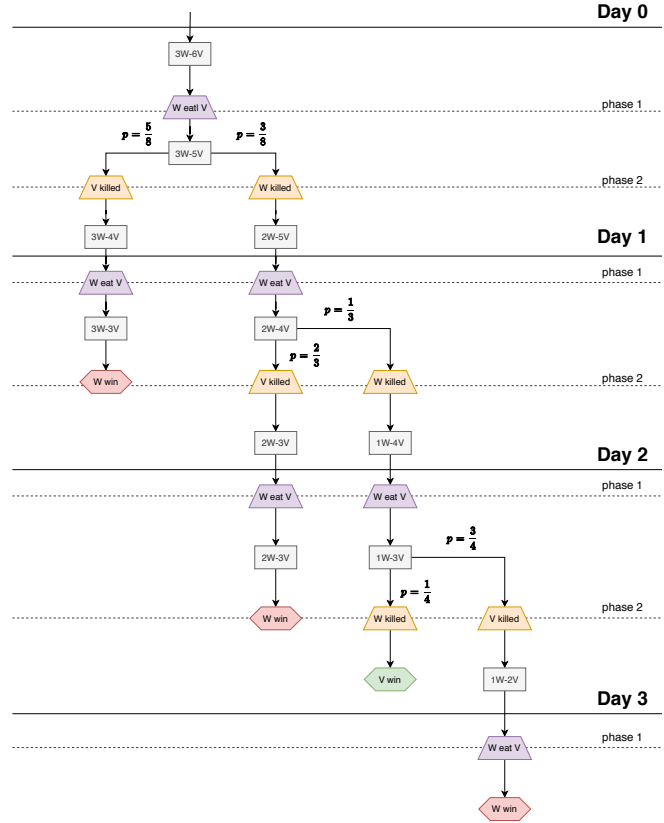}
    \caption{Expanded tree for a 9 players game with 3 wolves and 6 villagers}
    \label{fig:9P:expanded}
\end{figure}

An expanded tree, or tree of possibilities, shows every outcome of a particular process. At each possible intersection, a new branch is created associated with a probability $p$; in the context of the werewolf game, each branch represents an outcome where, during the execution phase, the player has killed either a villager or a werewolf.\\
It is worth noticing that in a random 9P environment, the villagers are most likely to lose. As can be seen in the expanded tree for the game, \ref{fig:9P:expanded}, the villagers either execute a werewolf during the first day, with probability $p=3/8$, or they lose. Moreover the only chance to win is achievable with probability:
$$
p=\frac{3}{8}\cdot \frac{1}{3}\cdot \frac{1}{4}=0.03125
$$
Hence they have a probability of $3.12\%$ to win with a complete random policy. 

\newpage
\subsection{AiWolf}
\label{app:aiwolf}

This section aims to provide some further details concerning the different implementation between the \emph{RLupus} framework and the \emph{AiWolf} one \cite{aiwolf}.

\subsubsection{Syntax protocol}
On their \href{http://aiwolf.org/en/protocol}{website} they define a specific protocol for the structure of sentences which is the combination of 6 words, i.e. unit of meaning:
\begin{enumerate}
    \item \emph{Subject}: an agent identifier which has a clear mapping with the \emph{Own Id} parameter specified in the previous Section \ref{app:obs_space}
    \item \emph{Target}: an agent identifier, again the parallelism is the same with the \emph{RLupus} setting.
    \item \emph{Role}: one of the 6 implemented roles. This kind of specification is not needed in our setting since only two roles are available.
    \item \emph{Species}: either human or werewolf.
    \item \emph{Verb}: a set of 15 valid verbs.
    \item \emph{Talk number}: a unique id for each sentence.
\end{enumerate}

These atomic information units can define 13 different type of sentences which can be broadly divided into 5 categories:
\begin{enumerate}
    \item Sentences expressing knowledge or intent.
    \item Sentences about ongoing actions. 
    \item Sentences about past actions and their results.
    \item Sentences that express dis/agreement.
    \item Sentences referring to the flow of the conversation.
\end{enumerate}

Finally, there are 8 types of operators which are used to frame sentences and express their relationship:
\begin{enumerate}
    \item Request operators which are directed towards the acquisition of new knowledge.
    \item Reasoning operators: e.g., because.
    \item Time indication operators.
    \item Logic operators such as: and, not, or, xor.
\end{enumerate}

\subsubsection{Comparison}
An initial parallelism can be traced between the number of words and the signal parameters \footnote{SL, SR in Section \ref{app:obs_space}}. Still, the main difference is that \emph{AiWolf} forces a prior meaning on the available words while \emph{RLupus} leaves them free to be used by the agents. On one hand, defining a grammar and a syntax beforehand allows for a highly interpretable language, but on the other, this constrains the agents to adapt to a syntax that could not be optimal. 

Moreover, since a sentence must be grammatically correct, some constraint enforcing this aspect must be applied to the agent through policy-shaping or on their output in a later phase. This further increases the complexity of the \emph{AiWolf} environment and define a solid discrepancy with the freedom expressed in the \emph{RLupus} one. 

Finally, in the RLupus framework, no direct relationship between sentences can be express by the agents since no unique id is defined for a sentence in the observation space.

\end{document}